\documentclass[aps,prb,superscriptaddress,preprint]{revtex4-1}

\usepackage{amsmath}
\usepackage[pdftex]{graphicx}
\usepackage{epstopdf}
\usepackage{xcolor}
\usepackage{booktabs}
\usepackage{relsize}
\usepackage{ifthen}

\def \bn{\begin{align}}
\def \en{\end{align}}
\def \be{\begin{equation}}
\def \ee{\end{equation}}
\def \bea{\begin{eqnarray}}
\def \eea{\end{eqnarray}}
\def \ba{\begin{array}}
\def \ea{\end{array}}

\def \av#1{{\langle#1\rangle}}

\def \w{{\omega}}

\def \yd{^\dagger}

\def \etal {{\it et al. }}

\usepackage{ulem}

\renewcommand{\epsilon}{\varepsilon}

\def \mysection#1{{\bf #1 }}

\newcommand{\forloop}[5][1]%
{%
\setcounter{#2}{#3}%
\ifthenelse{#4}%
	{%
	#5%
	\addtocounter{#2}{#1}%
	\forloop[#1]{#2}{\value{#2}}{#4}{#5}%
	}%
	{%
	}%
}%

\begin{document}

\title{Holographic Maps of Quasiparticle Interference}
\begin{abstract}

The analysis of Fourier-transformed scanning-tunneling-microscopy (STM) images with subatomic resolution is a common tool for studying properties of quasiparticle excitations in strongly correlated materials. While Fourier amplitudes are generally complex valued, earlier analysis mostly considered only their absolute values. Their complex phases were deemed random, and thus irrelevant, due to the unknown positions of impurities in the sample. Here we show how to factor out these random phases by analysing overlaps between Fourier amplitudes that differ by reciprocal lattice vectors. The resulting holographic maps provide important and previously-unknown information about the electronic structures of materials. When applied to superconducting cuprates, our method solves a long-standing puzzle of the dichotomy between equivalent wavevectors. We show  that $d$-wave Wannier functions of the conduction band provide a natural explanation for experimental results that were interpreted as evidence for competing unconventional charge modulations. Our work opens a new pathway to identify the nature of electronic states in STM measurements.

\end{abstract}

\author{Emanuele G. Dalla Torre}
\affiliation{Department of Physics, Bar Ilan University, Ramat Gan 5290002, Israel}
\author{Yang He}
\affiliation{Department of Physics, Harvard University, Cambridge, MA 02138, U.S.A.}
\author{Eugene Demler}
\affiliation{Department of Physics, Harvard University, Cambridge, MA 02138, U.S.A.}
\date{\today}
\maketitle

\mysection{Introduction}
Fourier-transformed scanning tunnelling spectroscopy\cite{sprunger97_giant} (FT-STS) is a powerful technique based on Fourier transforming the local density of states measured by a scanning tunnelling microscope (STM). The resulting maps deliver information about the scattering of quasiparticles in the conduction band, and can be employed to reconstruct their dispersion relation\cite{podolsky03,lee03,capriotti03_wave,polkovnikov03,pereg03_theory,markievicz04,hirschfeld04,bena2004quasiparticle}. In the last two decades, this technique has found important applications in the study of high-temperature-superconducting cuprates \cite{hoffman02A,mcelroy2003relating,hanaguri09,vishik2009momentum} and pnictides \cite{hanke12_probing}, semimetals \cite{vonau2005_evidence}, graphene \cite{rutter2007_scattering}, topological insulators \cite{roushan2009topological}, heavy fermions \cite{lee09_heavy,toldin2013disorder}, and more. Recent technological advances have provided STM images with a subatomic resolution, or equivalently FT-STS maps that include several Brillouin zones. In contrast to the naive expectation, the experimental signal is not periodic as a function of momentum: FT-STS maps at wavevectors that differ by a reciprocal lattice vector can sometimes be quantitatively or even qualitatively different\cite{chatterjee06_nondispersive,davis07,hudson08,yazdani10,fujita11}. 

To understand these discrepancies, Fujita \etal  \cite{fujita14} introduced a new ``phase sensitive'' analysis of the STM data, obtained by multiplying the real-space signal by specific atomic masks motivated by material-dependent considerations\cite{allais14_density,thomson2015charge}. It was later proposed \cite{yanghe15} to improve this analysis by directly shifting the FT-STS maps in momentum space. Here we generalize this approach and consider the overlap of FT-STS maps shifted by an arbitrary Bravais vector of the reciprocal lattice. This procedure allows us to recover the useful part of the phase information of the quasiparticle scattering, which can be employed to reconstruct the nature of the electronic states in the conduction band. We refer to the resulting maps as holographic maps, or $h$ maps, in analogy to optical holography, where both the intensity and the phase of the scattered waves are recorded. 

\begin{figure}[t]
\centering
\includegraphics[scale=0.4]{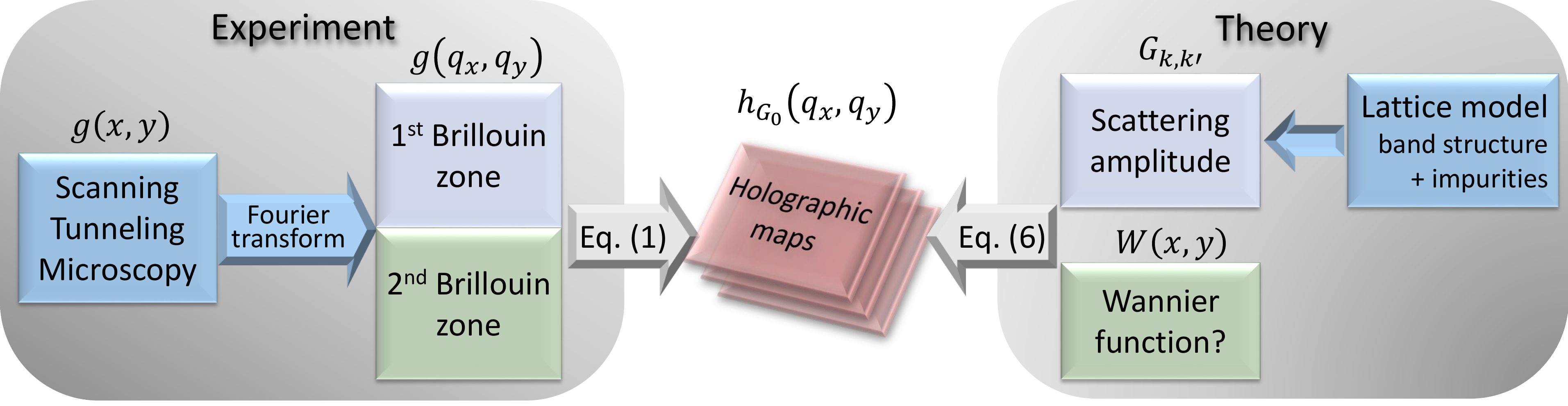}
\caption{{\bf Main steps and notations involved in the creation of holographic maps.} $h$ maps are generated independently from the experimental measurements and from a phenomenological lattice model (see text for details). The comparison between the $h$ maps obtained by the two methods is used to deduce the shape of the Wannier function.}
\label{fig:schematic2}
\end{figure}

Holographic maps can be easily generated from experimentally-measured FT-STS maps, by considering the product of the signal at equivalent wavevectors, i.e. by multiplying FT-STS maps from different Brillouin zones. The resulting $h$ maps deliver new information about properties of the quasiparticle scattering at the atomic level. These maps can be theoretically predicted by first computing the scattering amplitude from a one-band (or few-band) effective model, and then convoluting it with a trial Wannier function (see also Fig.~\ref{fig:schematic2}). The comparison between the experimental measurement and the theoretical calculation provides a stringent test on both the nature of scattering and the form of the Wannier function. 
The validity of the present approach is exemplified below for superconducting cuprates, where our analysis confirms the expected $d$-wave shape of the Wannier function. The experimental measurement of $h$ maps demonstrates that FT-STS maps of cuprates have a global $s$-wave rotational symmetry and are consistent with a model of Friedel oscillations around local impurities\cite{dallatorre13,dallatorre15}. Because the only major assumption of our work is the 
validity of Bloch theorem for conduction-band electrons, we expect our approach to find important applications in the study of strongly correlated materials. In particular, the present method may shed light on materials like heavy-fermions, where the nature of the band that is responsible for superconductivity is still debated \cite{allan2013imaging}.

\mysection{Properties of the holographic maps} FT-STS maps are obtained by the two-dimensional Fourier transform of the differential tunnelling conductance, $g({\bf r},V)=dI({\bf r})/dV$, measured in STM experiments. The $g$ maps defined by $g({\bf q},V) = \int d^2r~ e^{i {\bf q\cdot r}} g({\bf r},V)$ are generically complex valued. 
In disordered samples their phase depends on the position of the scatterers. (This is a simple consequence of the fundamental property of Fourier transformations, stating that a shift in real space corresponds to a phase rotation in momentum space.) If a single impurity is present \cite{davis99,davis01} this phase is given by $e^{i {\bf q \cdot r_0}}$, where ${\bf r_0}$ is the position of the impurity, and can be factored out by choosing the axis origin  in correspondence to the centre of the impurity (${\bf r}_0=0$). In contrast, when analysing large samples, several scatterers should be taken into account. In this case, the phase of the $g$ map depends on the random configuration of the disorder\cite{zeljkovic12,hamidian12_picometer} and is therefore often assumed to provide little information about the material. One possible solution is to consider the auto-correlations of the STM signal in real space, which imply an ensemble average over the position of the impurities. This method was employed for example by Ref.~[\onlinecite{richardella2010_visualizing}] to study electronic states  close to the superconducting-insulator phase transition.

%
Here we instead propose to work in momentum space and to consider the overlaps of $g$ maps shifted by a generic reciprocal lattice vector ${\bf G}$. Specifically, we define the holographic maps
\be h_{\bf G}({\bf q},\omega) = g({\bf q},\omega) g^*({\bf q}+{\bf G},\omega)\label{eq:Hmap}\;.\ee 
As will be shown below, under realistic assumptions both the absolute value and the phase of $h_{\bf G}({\bf q},\omega)$ do not depend on the random position of the impurities on the lattice. In the case of ${\bf G=0}$, Eq.~(\ref{eq:Hmap}) simply reduces to the $|g({\bf q},\omega)|^2$, confirming the known fact that the absolute value of $g$ maps does not depend on the position of the impurities. For all other Brillouin vectors ${\bf G}\neq 0$, Eq.~(\ref{eq:Hmap}) delivers additional information that was previously hidden in the $g$ maps.

To clarify the effects of disorder on Fourier-transformed maps, it is useful to consider $N$ identical randomly-distributed weak scatterers on a lattice. To the lowest order in perturbation theory, the resulting $g$ map is $g({\bf q},\omega) = \sum_i e^{i{\bf q\cdot r}_i} g_0({\bf q},\omega)$, where ${\bf r}_i$ are the positions of the scatterer and $g_0({\bf q},\omega)$ is the map generated by a single scatterer. Performing an ensemble average over the position of the scatterers we obtain $\overline{g({\bf q},\omega)}=0$ and $\overline{ |g({\bf q},\omega)|^2}=\sum_{i,j} \overline{e^{i {\bf q \cdot (r_i-r_j)}}} |g_0({\bf q},\omega)|^2 =N|g_0({\bf q},\omega)|^2 $, where $N$ is the number of impurities and we used $\overline{e^{i {\bf q \cdot (r_i-r_j)}}}=\delta_{r_i,r_j}$. As expected, we find that although $g({\bf q},\omega)$ averages to zero, its absolute value remains finite
. In principle, a similar argument would suggest that ensemble-averaged $h$ maps should vanish as well. This is however not the case if all the scatterers are physically equivalent, such as in the case of identical chemical substitutions, all located in the same positions of the unit cell. For this ensemble ${\bf G} \cdot {\bf r}_i = \phi_0$ is constant and 

\begin{align}  \overline{h_{\bf G}({\bf q},\w)}&=\sum_{i,j} \overline{e^{i {\bf q \cdot (r_i-r_j) + i G \cdot {r_j}}}} g_0({\bf q},\omega)g^*_0({\bf q+G},\omega) = N e^{i\phi_0} g_0({\bf q},\omega)g^*_0({\bf q+G},\omega)\; .
\end{align}
We find that $h$ maps are invariant under ensemble average, up to an overall wavevector-independent phase factor, $e^{i\phi_0}$. (As explained in the Methods section, this phase factor can be removed by an appropriate choice of the origin of the Fourier transform.)
In actual materials several types of scatterers are often present, for example as a consequence of distinct dopants, oxygen vacancies, or pinned vortices. For simplicity in this paper we focus on the case of one dominant type of scatterers, relevant to superconducting cuprates \cite{zeljkovic12,dallatorre15}. Extension to several sources of disorder is straightforward.

\begin{figure}[t]
\includegraphics[scale=1]{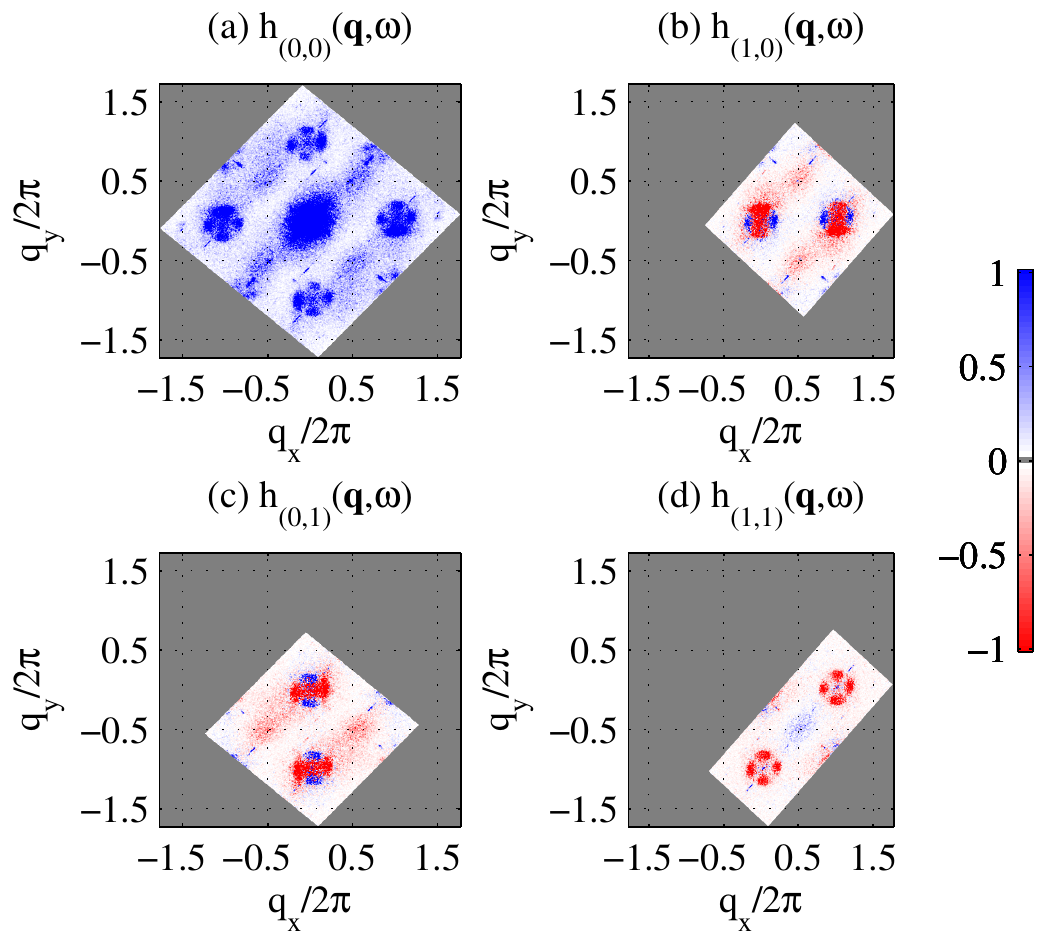}
\caption{{\bf Experimentally-measured holographic maps} for an optimally doped sample of Bi2201 with critical temperature $T_c=35$K, at voltage $V=40$meV. (a) When ${\bf G}=(0,0)$, the $h$ map simply corresponds to the square of the absolute value of the FT-STS map, which has been previously reported in the literature\cite{yanghe15}; (b-c) For ${\bf G}=(1,0)\times (2\pi/a)$ and ${\bf G}=(0,1)\times (2\pi/a)$ the $h$ map displays star-shaped patterns with d-wave symmetry, which demonstrate the necessity of non-trivial Wannier functions to interpret the data (see text for details); (d) For ${\bf G}=(1,1)\times (2\pi/a)$ the $h$ map is negative, in agreement with the $d$-form factor proposed by Refs.~[\onlinecite{allais14_density,thomson2015charge,fujita14,hamidian2015atomic,hamidian15_magnetic,machida2015bipartite,yanghe15}]. Grey areas refer to wavevectors that are beyond the present experimental resolution.}
\label{fig:dwave_exp}
\end{figure}



\begin{figure}[t!]
\begin{tabular}{c c c}
\includegraphics[scale=0.5]{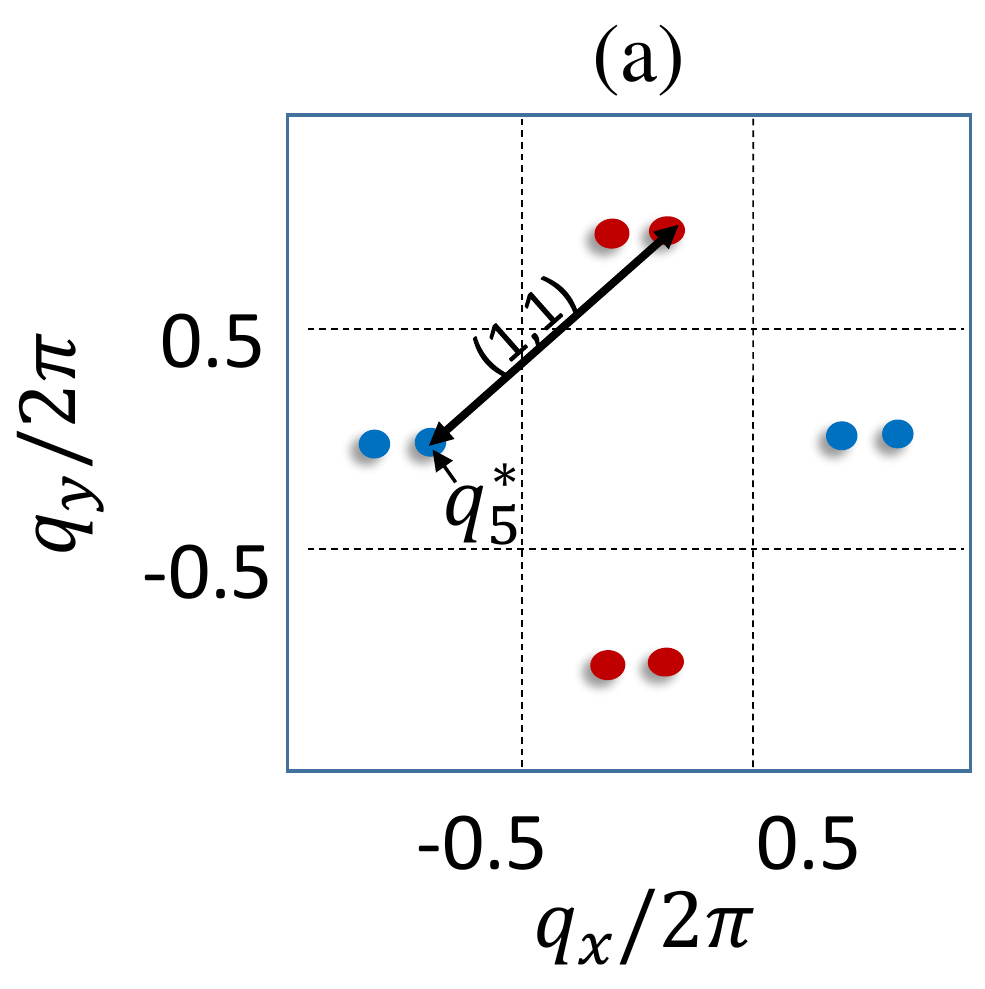}&
\includegraphics[scale=0.5]{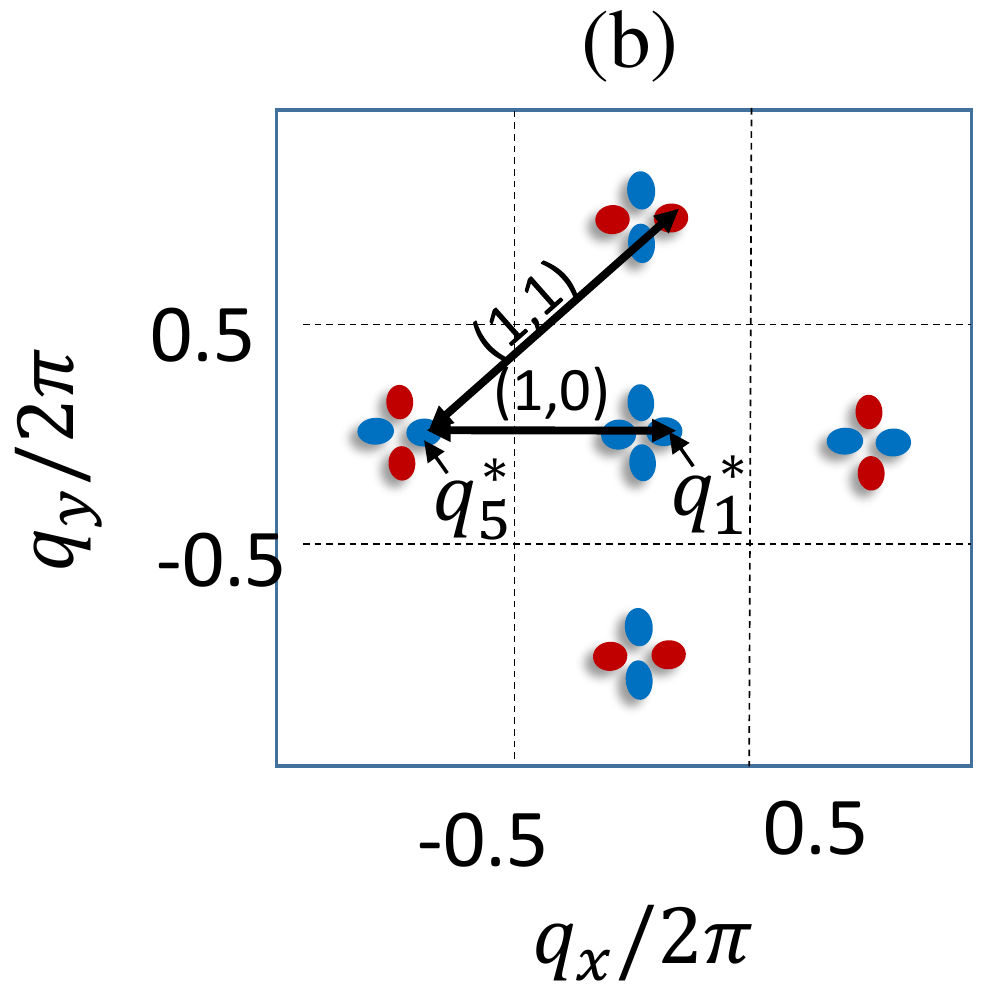}&
\includegraphics[scale=1.0]{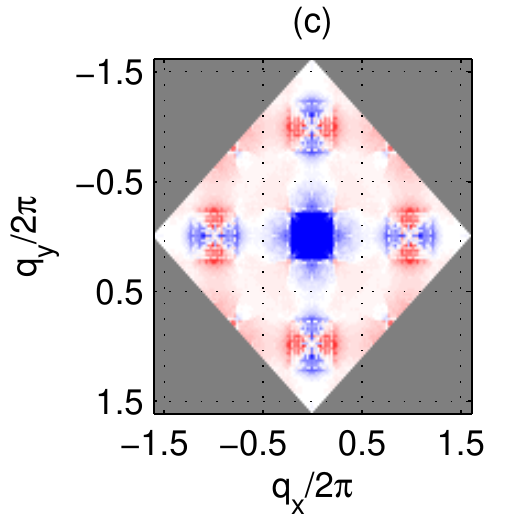}
\end{tabular}
\caption{{\bf Schematic and theoretical $\bf g$ maps} (a) Schematic plot of the $d$-form factor proposed by Refs.~[\onlinecite{allais14_density,thomson2015charge,fujita14,hamidian2015atomic,hamidian15_magnetic,machida2015bipartite,yanghe15}], implying that $q_5^*$ peaks that differ by (1,1) have opposite signs. This finding is consistent with the present analysis, which additionally shows a coherence between the $q_1^*$ and $q_5^*$ peaks that differ by (1,0). (b) Schematic plot of the $g$ map deduced from the present analysis, characterized by a global $s$-wave rotational symmetry. (c) Numerically computed $g$ map at $V=40$meV, showing the same symmetry as subplot (b). These $g$ maps assume an isolated scatterer at the axis origin and cannot be directly compared to experiments due to the random position of the impurities in actual experimental samples. The correctness of our interpretation is however demonstrated by the symmetry of the corresponding holographic maps.}
\label{fig:schematic-gmap}
\end{figure}

\mysection{Experimental measurement of $\bf h$ maps} To demonstrate the usefulness of the present approach, we consider the specific case of Bi$_2$Sr$_{2−x}$La$_x$CuO$_{6+\delta}$ (Bi2201). Experimentally-measured $h$ maps of this material are shown in Fig.~\ref{fig:dwave_exp}. (See Methods section for details and Fig.~\ref{fig:dwave_UD32K} of the SI for a second sample of the same material, at a different doping.) Subfigure (a) simply represents the square of the absolute value of the $g$ map. It displays a broad peak around ${\bf q}=0$, and four star-shaped patters at the four Bragg peaks ${\bf G}_0=(\pm1,0)$ and  $(0,\pm1)$. (For brevity we express all wavevectors in units of $2\pi/a$.) Each of these patters include four distinct satellites, located respectively at $ {\bf q}_5^* = {\bf G}_0\pm(\delta,0)$ and ${\bf G}_0\pm(0,\delta)$, where $\delta\approx0.2$ is material and doping dependent \cite{hudson08}. A closer inspection of the data \cite{hudson08} reveals pronounced shoulders in the central Brillouin zone as well, at wavevectors ${\bf q}^*_1 = (\pm \delta,0)$ and $(0,\pm \delta)$. These peaks are commonly interpreted \cite{hoffman02B,kapitulnik03,davis04,yazdani04,yanghe13} as evidence of a static (checkerboard?) modulation with wavevector $\delta$. The same type of incommensurate short-ranged order was recently observed in X-ray scattering experiments \cite{ghiringhelli12,chang12} and is currently the focus of an intense theoretical and experimental investigation\cite{damascelli2015new,grissonnanche2015onset,gerber15_three,Julien20112015}. For a long time, the observed difference between the visibility of the peaks at $q_1^*$ and $q_5^*$ has been regarded as an unsolved puzzle \cite{chatterjee06_nondispersive,davis07,hudson08,yazdani10,fujita11}. To better understand the nature of these peaks,  Fujita \etal \cite{fujita14} introduced a real-space masking procedure, which revealed that the $q_5^*$ peaks are characterized by a $d$-form factor, in agreement with theoretical predictions of Sachdev and collaborators\cite{allais14_density,thomson2015charge} (see Fig.~\ref{fig:schematic-gmap}(a)). This finding is confirmed by the $h_{(1,1)}$ map shown in Fig.~\ref{fig:dwave_exp}(d). This map is mainly negative indicating that the $q_5^*$ peaks that differ by (1,1) have opposite signs. Here we show that a non-trivial Wannier function offers a natural explanation for these findings, and is endorsed by the new  phase information offered by the $h$ maps.

In a nutshell, $h$ maps allow us to determine the phase coherence between the signal at $q_1^*$ and $q_5^*$. These peaks differ by the inverse lattice vector ${\bf G}_0=(1,0)$ and their overlap is encoded in $h_{(1,0)}$, reproduced in Fig.~\ref{fig:dwave_exp}(b). As expected, this map shows two identical patterns, respectively centred around $(0,0)$ and $(1,0)$. Each pattern includes four satellites, which originate from the overlaps between the four inequivalent ${\bf q}_1^*$ peaks and their corresponding ${\bf q}_5^*={\bf q}_1^*+{\bf G}_0$ wave-vectors. Remarkably, we find that these overlap patterns have a local $d$-wave symmetry. This finding indicates that either  the $q_1^*$  pattern has a $s$-wave symmetry and each of the $q_5^*$ patterns has a {\it local} $d$-wave symmetry, or vice versa. The former interpretation is consistent with the observed intensity of the FT-STS signal and explains the different visibility of the peaks. Because the $g$ map in the central Brillouin zone is $s$-wave symmetric, its intensity is roughly rotational invariant and partially hides the $q_1^*$ peaks. In contrast, the $g$ map in the first Brillouin zone has a local $d$-wave symmetry and vanishes along the two diagonals, accentuating  the four $q_5^*$ peaks. Combining this information with the above-mentioned $d$-form factor  (i.e. the negative sign of the correlations in $h_{(1,1)}$), we obtain the $g$ map  schematically shown in Fig.~\ref{fig:schematic-gmap}(b). Remarkably, this map has a global $s$-wave symmetry (i.e. it is invariant under rotations of 90 degrees).

\mysection{Theoretical calculation of $\bf h$ maps} Wannier functions offer a natural explanation for the inequivalence of FT-STS maps in distinct Brillouin zones, and in particular  for the dichotomy between $q_1^*$ and $q_5^*$. The effects of Wannier functions on $g$ maps were first theoretically considered by Podolsky \etal \cite{podolsky03}, and later applied to the study of actual materials in Refs.~[\onlinecite{dallatorre13,choubey14,kreisel15,dallatorre15}]. One fundamental difficulty with this approach stems from the observation that Wannier functions are not uniquely defined, leading to an unexpected arbitrariness of the theoretical description of $h$ maps. In the Supplementary Information (SI) we solve this problem, by employing a Bloch-function approach. This method allows us to explicitly specify the Wannier function to be used, and avoid the phase problem that one encounters in their usual definition. Remarkably, the final expression has the same form as the equations used in Refs. [\onlinecite{podolsky03,dallatorre13,choubey14,kreisel15,dallatorre15}]

\vspace{-0.7cm}
\begin{align} 
g({\bf q},V) &= {\textrm i}\sum_{ \bf k} W({\bf k})W({\bf k+q}) G_{\bf \bar{k}, \overline{k+q}}(\omega)\;. \label{eq:gWWG}
\end{align}
Here $\bf \bar k \equiv k~ {\rm mod}~ G$ is restricted to the first Brillouin zone, while the sum over ${\bf k}$ runs over all wavevectors; $G_{\bf k,k'}(\omega)$ is the retarded Green's function of the conduction band, and corresponds to the scattering amplitude of quasiparticles from momentum ${\bf k}$ to ${\bf k'}$.

\begin{figure}[t!]
\includegraphics[scale=1]{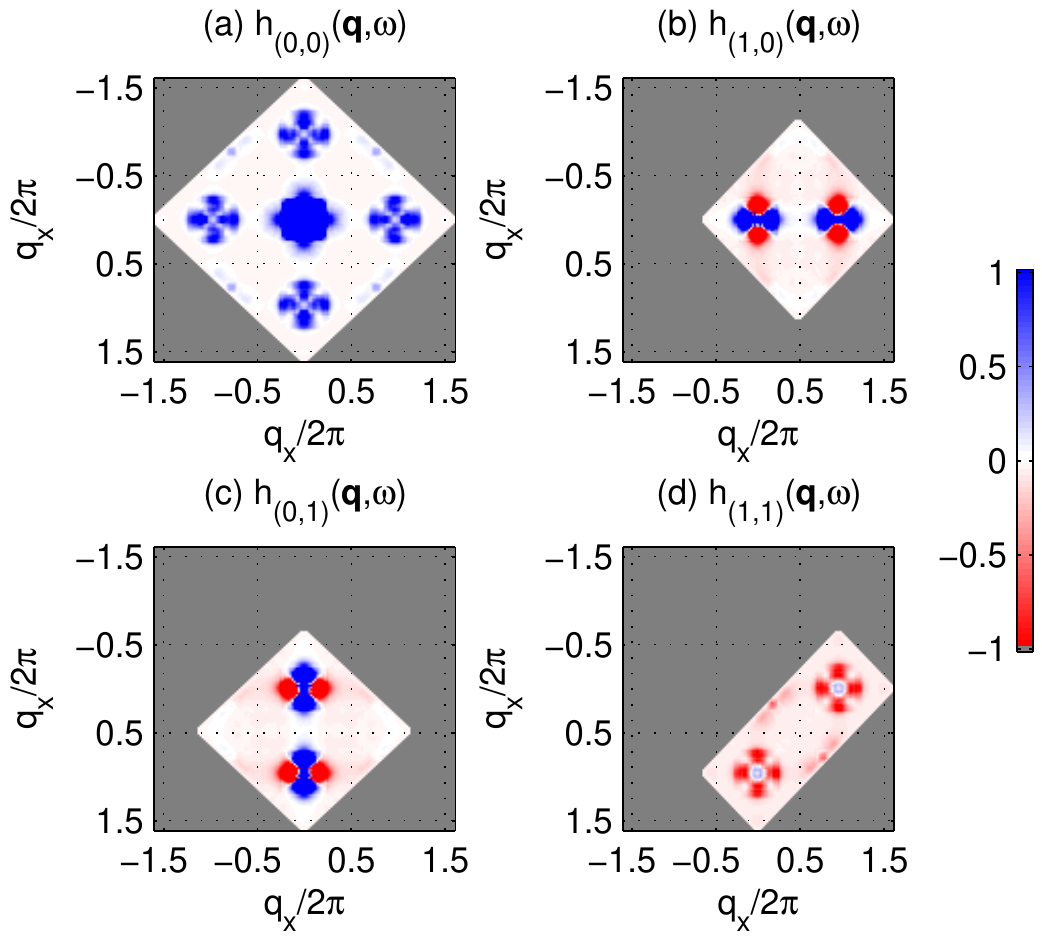}
\caption{{\bf Theoretical holographic maps} for four reciprocal lattice vectors, generated from the $g$ map depicted in Fig.~\ref{fig:schematic-gmap}(c). All four $h$ maps precisely reproduce the experimental results presented in Fig.~\ref{fig:dwave_exp}, including in particular the different symmetries and visibilities of the $q_1^*$ and $q_5^*$ peaks.}
\label{fig:dwave_th}
\end{figure}

To generate theoretical FT-STM maps we need to specify a lattice model for $ G_{{\bf \bar{\bf k}},{\overline{\bf k+q}}}$. Following our earlier findings\cite{dallatorre13,dallatorre15} we claim that the checkerboard-type short-range order observed in STM experiments simply corresponds to Friedel oscillations around local impurities, rather than to a competing charge-density-wave (CDW) order. We specifically consider the scattering of quasiparticles with finite lifetime from a local modulation of the pairing gap. This type of modulations can be induced by static disorder, or in the presence of an applied magnetic field, by the core of a pinned vortex\cite{wu11,wu13,dallatorre13}.  The resulting $g$ map is presented in Fig.~\ref{fig:schematic-gmap}(c) (see Methods section for details of the calculation). Note that this map has a {\it global} $s$-wave rotational symmetry and its structure coincides with the schematic plot inferred from the experiments in subplot (b). As explained above, in actual experiments, $g$ maps are multiplied by the random phase $e^{i{\bf q}\cdot {\bf r_i}}$, preventing a direct observation of the rotational symmetry. This random phase can be factored out by computing the $h$ maps. The results of our theoretical calculations are shown in Fig.~\ref{fig:dwave_th} and accurately reproduce all the details of the experimental findings (Fig~\ref{fig:dwave_exp}). In particular, our theoretical calculations account for the experimentally-observed symmetry of the $h_{(1,0)}$ maps and the dichotomy between the weaker features at $q_1^*$ and the sharper peaks at $q_5^*$ .

\begin{figure}
\centering
\begin{tabular}{c c c}
(a) $x^2 - y^2$  & (b) $x^2 + y^2$ & (c) $|x^2 - y^2|$ \\
\includegraphics[scale=1.0]{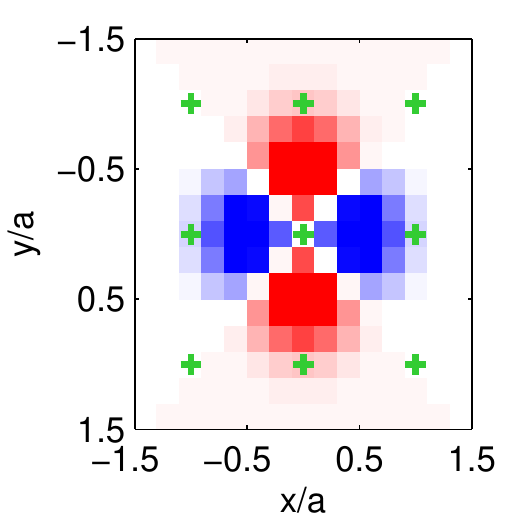} &
\includegraphics[scale=1.0]{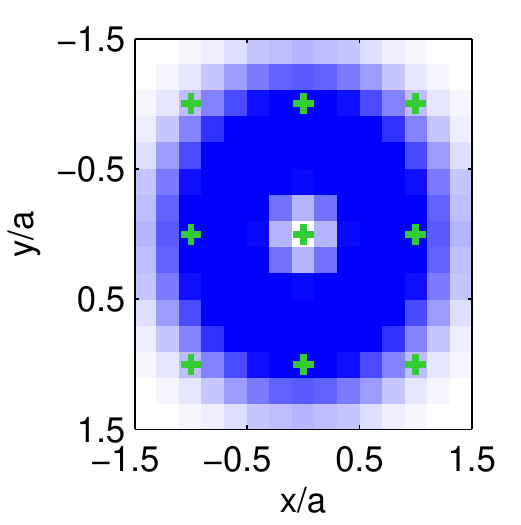} &
\includegraphics[scale=1.0]{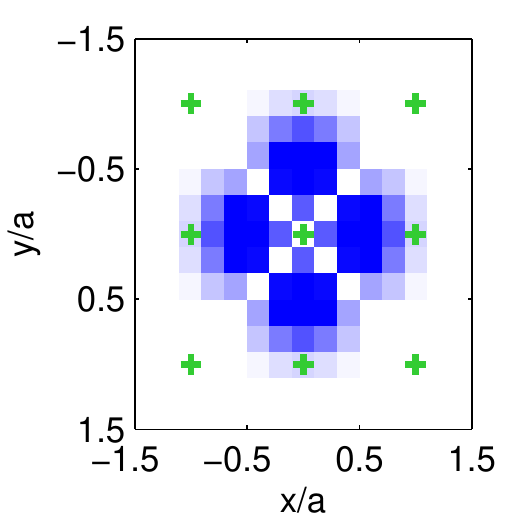}
\end{tabular}
\caption{{\bf Trial Wannier functions} (a)  $W(x,y)=(x^2-y^2)e^{-(x^2+y^2)/2\sigma^2}$, (b) $W(x,y)=(x^2+y^2)e^{-(x^2+y^2)/2\sigma^2}$, and (c)
$W(x,y)=|x^2-y^2|e^{-(x^2+y^2)/2\sigma^2}$. Here $\sigma=0.3a$ and the green crosses represent Copper atoms. The Wannier function   (a) corresponds to Eq.~(\ref{eq:Wx}) and was used to generate the numerical maps of Figs.~ \ref{fig:schematic-gmap} and \ref{fig:dwave_th}. The Wannier function (b) leads to $h$ maps that are inconsistent with the experimental measurements, while the Wannier function (c) generates maps that are similar to those of (a) (see \ref{app:wannier}). These three Wannier functions highlight that holographic maps are sensitive to the shape of the absolute of the Wannier function and not to its phase symmetry.  Interestingly, the Wannier functions (a) and (c) are peaked in correspondence to the in-plane oxygen atoms (see \ref{app:wannier}). }
\label{fig:wf}
\end{figure}

The theoretically predictions shown in Fig.~\ref{fig:dwave_th} were obtained using a Wannier function with $d$-wave symmetry and Gaussian envelope
\be
W({\bf x})= (x^2-y^2) e^{-(x^2+y^2)/(2\sigma^2)}\label{eq:Wx}
\ee
where $\sigma$ sets the global size of the Wannier function. 
In our calculations this parameter was used as the {\it only free fitting parameter}. The best agreement between theory and experiment was obtained for $\sigma=0.3a$, where $a$ is the unit cell length, leading to the Wannier function plotted in Fig.~\ref{fig:wf}(a). Interestingly, this wavefunction is significantly narrower than the one predicted by  the first-principle calculations of Kreisel \etal \cite{kreisel15}, for which $\sigma\approx 0.7a$. An intuitive argument for the relation between this Wannier function and the resulting FT-STS maps is provided in \ref{app:wannier}.
%
%
To further highlight the fundamental role of the Wannier function on the FT-STS maps, in Figs.~\ref{fig:theory1}-\ref{fig:theory4} of the SI we present calculations for alternative functions with different size and symmetry. In particular, we consider two families of extended $s$-wave Wannier functions, respectively proportional to $x^2+y^2$ and $|x^2-y^2|$, (see Fig.~\ref{eq:Wx}(b) and (c)). While both functions have the same symmetry under rotations by 90 degrees, the resulting $h$ maps are very different and only the latter is consistent with the experimental results. On the other hand, we observe that the $h$ maps generated by $W\sim x^2-y^2$  and $W\sim |x^2-y^2|$ are very similar. These results demonstrate that FT-STS maps  depend on the shape of the absolute value of the Wannier function and not on its phase symmetry. In our model, FT-STS maps are always $s$-wave  symmetric, irrespective of the  symmetry of the Wannier function. To discern between $d$-wave ($W\sim x^2-y^2$) and $s$-wave ($W\sim |x^2-y^2|$) Wannier functions it is therefore necessary to invoke additional arguments, such as the physical requirement of the Wannier function to be analytic (to avoid large kinetic-energy costs) or reasoning based on the relevant atomic orbital (see for example Zhang\&Rice\cite{zhang88}). In our case, both arguments support the $d$-wave Wannier function of Eq.~(\ref{eq:Wx}) and Fig.~\ref{fig:wf}(a).

\mysection{Discussion}
To summarize, we introduced holographic maps ($h$ maps) as a new tool to analyse Fourier-transformed STM images. These complex maps cancel random phase factors originating from the positions of impurities. We show that when the impurities are predominantly of the same type, the $h$ maps reveal intrinsic sign changes, which were previously unnoticed in the commonly used $g$ maps. This information can be exploited to derive the shape of Wannier functions of conduction-band electrons in strongly-correlated materials. In the specific case of superconducting cuprates, our method unveils a Wannier function with a $d$-wave shape and an extension of about one unit cell. This result is consistent with first-principle calculations\cite{kreisel15}, although the spatial extent is significantly smaller than previously predicted. Our analysis further explains the large visibility of the four satellites that were observed in FT-STS maps of the first Brillouin zone. Previous studies interpreted these peaks as evidence of a competing charge density wave \cite{hoffman02B,kapitulnik03,davis04,yazdani04,chatterjee06_nondispersive,davis07,hudson08,yazdani10,fujita11,yanghe13,fujita14,hamidian2015atomic,machida2015bipartite,yanghe15}. We offer a simpler interpretation in terms of Friedel oscillations and non-trivial Wannier functions, and  show that the FT-STS maps have a {\it global} $s$-wave rotational symmetry. Applications to other strongly-correlated materials are readily viable and may help to solve long-standing debates about the nature of conduction bands in these materials.

\section*{Methods}
In this section we provide some technical details of the procedures used to generate Figs.~\ref{fig:dwave_exp}-\ref{fig:dwave_th}. 
The analysis of the {\it experimental} data consists of four main steps: (i) The experimental STM data at a fixed voltage is first Fourier transformed to generate a $g$ map. As explained in the main text, these maps are characterized by a random momentum-dependent phase, related to the positions of the impurities. 
(ii) Next, the $g$ map is multiplied by a planar-wave $e^{i(q_x x_0 + q_y y_0)}$, where $x_0$ and $y_0$ are chosen to maximize the visibility of resulting $h$ maps. This step corresponds to a specific choice for the origin of the axes used to perform the Fourier transform. (iii) The $h$ maps (\ref{eq:Hmap}) are computed by multiplying shifted $g$ maps with the complex conjugate of the unshifted map. (iv) The real part of the $h$ maps is plotted on a colorplot. 
The entire procedure would be valid even if step (ii) is skipped. In this case however, each $h$ map would be multiplied by a global overall phase factor $e^{i\phi_0}=e^{i {\bf G}\cdot {\bf r}_0}$, with arbitrary ${\bf r}_0$. In the case of Fig.~\ref{fig:dwave_exp}(b) and (c) the relevant factors, $e^{i\phi_a} = e^{i x_0/a}$ and $e^{i\phi_b}=e^{i y_0/a}$, can lead to a random flip of the overall sign, exchanging red and blue colours. Note however that the phase factor associated with Fig.~\ref{fig:dwave_exp}(d) is $e^{i\phi_d}=e^{i (x_0+y_0)/a}=e^{i\phi_b+i\phi_c}$. Thus, although the signs of Fig. \ref{fig:dwave_exp}(b-d) are arbitrary, their product is uniquely determined. 

In our {\it theoretical} calculations we computed the scattering amplitude $G_{\bf k,k'}$ within the Born approximation (first order perturbation theory in the scatterer strength). Specifically, we used $G_{\bf k,k'}=G_0({\bf k'})\Big(T({\bf k}+T({\bf k'})\Big)G_0({\bf k'})$. Here $G_0({\bf k'})$ is the Green's function of a paired state in Nambu space and $T({\bf k})=(\cos(k_x)-\cos(k_y))\sigma_x$, where $\sigma_x$ is a Pauli matrix, models a local modulation of the pairing gap. (See Ref.~[\onlinecite{dallatorre15}] for details of the calculations and Fig.~\ref{fig:dwave_v1} for the effects of local modulations of the chemical potential.) The phenomenological parameters used are (i) The band structure of Ref.~[\onlinecite{campuzano95}]; (ii) A chemical potential leading to the Luttinger-count doping $p=0.2$; (iii) A pairing gap $\Delta_0=20$meV; (iv) A quasiparticle lifetime $\Gamma=8$meV. The sum over ${\bf k}$ in Eq.~(\ref{eq:gWWG}) was performed by evaluating the summand over a grid of 200$\times$200 points covering the square included between $(-5\pi,-5\pi)$ and $(5\pi,5\pi)$. We verified that the numerical results are stable to small changes of the model, such as details of the phenomenological band-structures, and small changes of the doping and of the pairing gap \cite{dallatorre15}. In contrast, the specific choice of the impurity has important consequences for the shape of the $h$ maps (see \ref{app:scatterer}), highlighting the non-trivial relation between the source of disorder and the shape of the Friedel oscillations. Both the theoretical and experimental $h$ maps shown in this manuscript refer to the voltage $V=40$meV. See also \ref{app:energy} for some considerations concerning the voltage dependence of the signal. 


\newpage
\section*{Acknowledgment} We are grateful to Mike Boyer, Kamalesh Chatterjee, Eric Hudson, and Doug Wise for giving us access to their unpublished experimental data. We acknowledge useful discussions with Seamus Davis, Jennifer Hoffman, Eric Hudson, Robert Markiewicz, and Subir Sachdev. ED acknowledges support from
Harvard-MIT CUA, NSF Grant No. DMR-1308435, AFOSR Quantum Simulation MURI, the ARO-MURI on Atomtronics, ARO MURI Quism program, Dr.~Max R\"ossler, the Walter Haefner Foundation, and the ETH Foundation. EGDT was supported by the Israel Science Foundation (grant No.1542/14)

\newpage

\renewcommand{\theequation}{S\arabic{equation}}   
\setcounter{equation}{0}

\renewcommand{\thefigure}{S\arabic{figure}}   
\setcounter{figure}{0}

\renewcommand{\thesection}{SI-\arabic{section}}   
\setcounter{section}{0}

\renewcommand{\thetable}{S\arabic{table}}   
\setcounter{table}{0}

\section*{Supplementary Information}

\section{Effects of Wannier functions on $\bf g$ maps}

In this section we provide a derivation of Eq.(\ref{eq:gWWG}) based on Bloch theorem. In contrast to earlier attempts \cite{podolsky03,dallatorre13,choubey14,kreisel15,dallatorre15} our derivation circumvents the ambiguity of the definition of Wannier functions. 
The tunnelling conductance measured in STM experiments is proportional to the local density of states $g({\bf r},V)={\rm Im}[G^R({\bf r},\omega)]$, where the retarded Green's function $G^R({\bf r},\omega)={\mathrm i}\int_0^\infty~e^{-i V t} \lbrace \psi\yd({\bf r},t),\psi({\bf r},0)\rbrace$.
In a periodic lattice Bloch theorem states that $\psi({\bf r},t)=\sum_k e^{i k r} \phi_{\bf k}({\bf r}) \psi_k$, where $\psi_{\bf k}$ annihilates a quasiparticle with momentum ${\bf k}$, the sum of ${\bf k}$ runs over the first Brillouin zone only, and $\phi_{\bf k}({\bf r})$ are Bloch wavefunctions. These functions are defined to be periodic over the unit cell, or $\phi_{\bf k}({\bf r}+{\bf  R})=\phi_{\bf k}({\bf r})$, where ${\bf R}$ is a real-space Bravais vector. Using this definition we obtain:

\begin{align} 
g({\bf q},V) &= {\rm Im} \int d^d r \sum_{\bf k, k'} e^{i {\bf q\cdot r}} e^{i {\bf (k-k')\cdot r}} \phi_{\bf k}({\bf r})\phi^*_{\bf k'}({\bf r}) G_{\bf k,k'}(\omega) \;,
\end{align}
where we introduced the Green's function $G_{\bf k,k'}(\omega)={\textrm i}\int_{0}^{\infty} dt~e^{-i\omega t}\av{[\psi_{\bf k}(t),\psi_{\bf k'}\yd(0)]}$. Bloch wavefunctions are periodic over a unit cell and can be expanded as a Fourier series $\phi_{\bf k}({\bf r}) = \sum_{\bf G}\phi_{\bf k}({\bf G}) e^{i{\bf G \cdot r}}$, where ${\bf G}$ runs over the inverse lattice vectors. We then obtain
\be
\ba{l}g({\bf q},V) = {\textrm i}\mathlarger{\sum} \phi_{\bf k}({\bf G})\phi^*_{\bf k'}({\bf G'}) G_{\bf k,k'}(\omega)\;, \\{ \ba{c}~~~~~~\scriptstyle \bf k, k',G,G'\\~~~~~~~\scriptstyle{\bf k-k'+G-G'+q=0}\ea} \ea\label{eq:FFG}
\ee
where the sum over ${\bf k}$ and ${\bf k'}$ runs over the first Brillouin zone only. Eq.~\ref{eq:FFG} can be formally simplified by introducing the Fourier-transformed Wannier functions $W({\bf k}) =\phi_{\bf {\bar k}}({\bf G})$, where we defined $\bf \bar k = k~ {\rm mod}~ G$, and ${\bf k}$ is arbitrarily large.  Substituting this definition into Eq.~(\ref{eq:FFG}) we obtain Eq.~(\ref{eq:gWWG}).
 
\section{Holographic maps of an underdoped sample}
\begin{figure}[h!]
\includegraphics{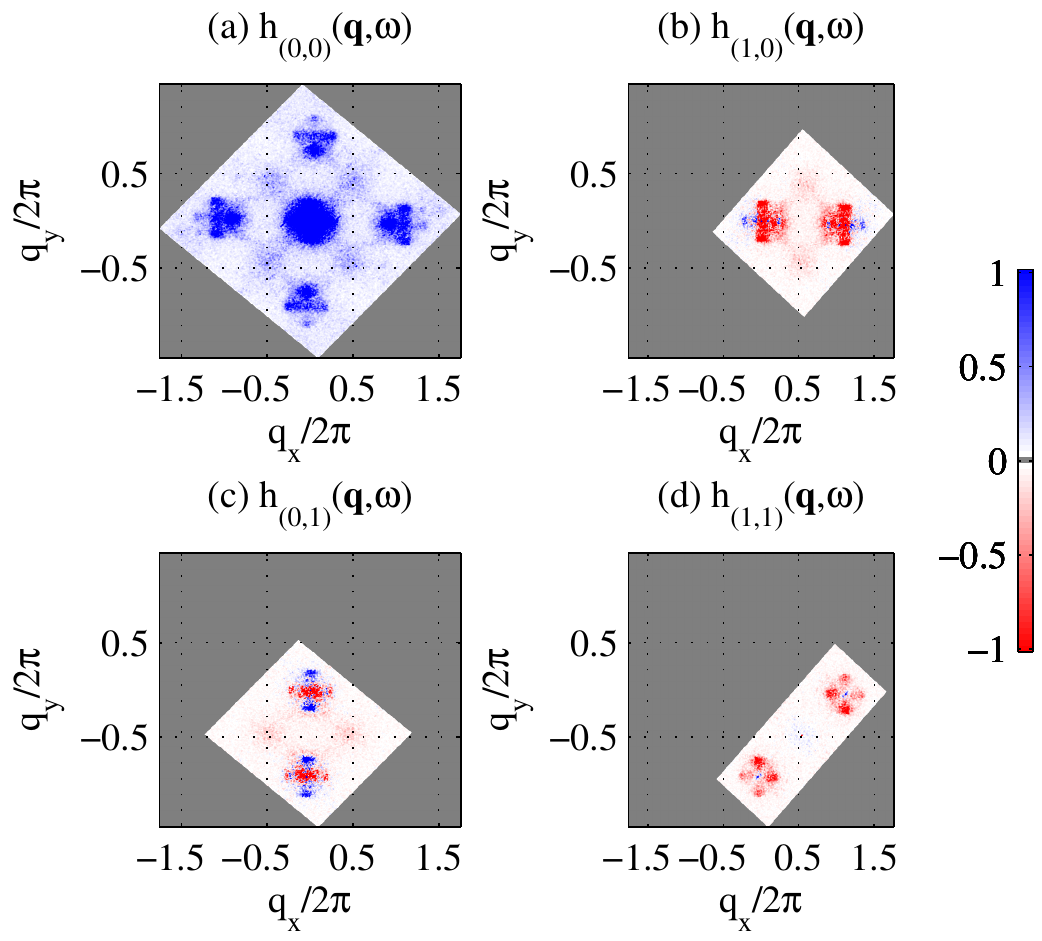}
\caption{{\bf Experimentally-measured holographic maps} for an underdoped sample of Bi2201 with critical temperature $T_c=32K$, at $V=30$meV. The $h$ maps of this sample display the same symmetries as those of the optimally-doped sample shown in Fig.~\ref{fig:dwave_exp}.}
\label{fig:dwave_UD32K}
\end{figure}

See Fig.~\ref{fig:dwave_UD32K}.

\begin{figure}
\includegraphics{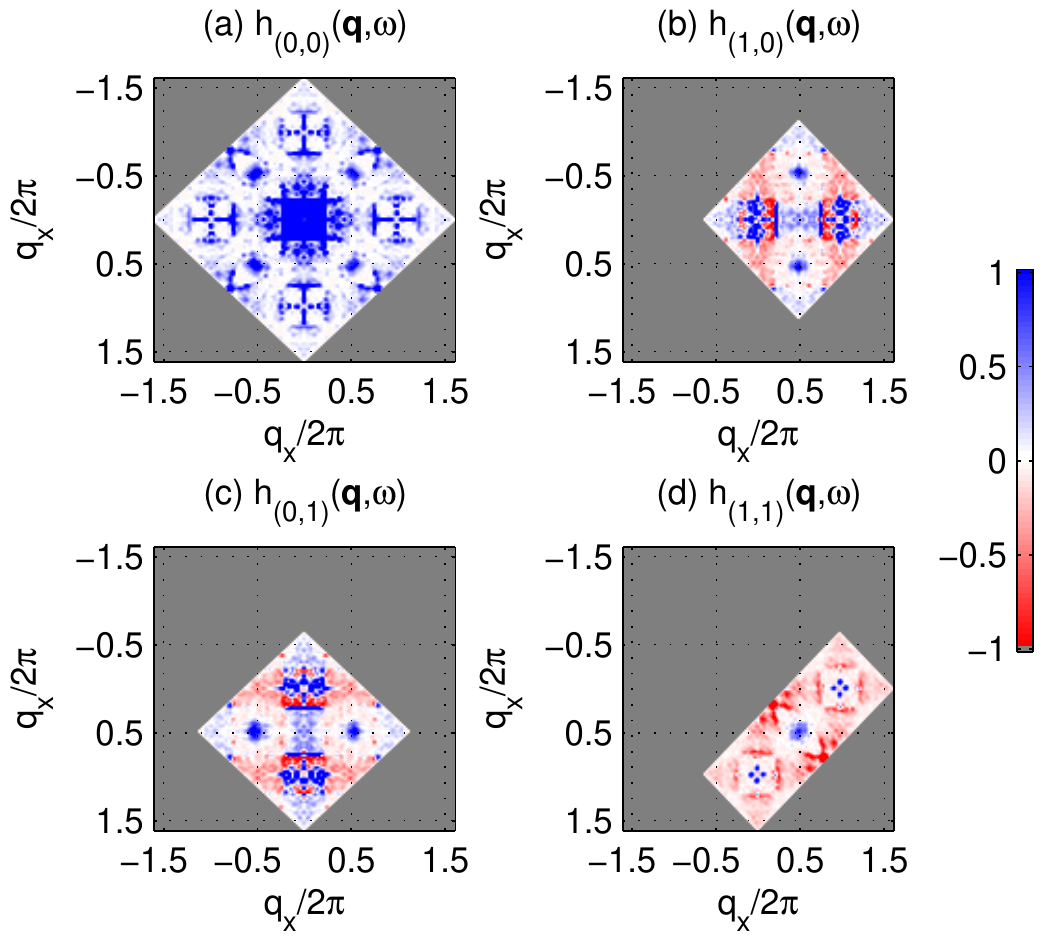}
\includegraphics{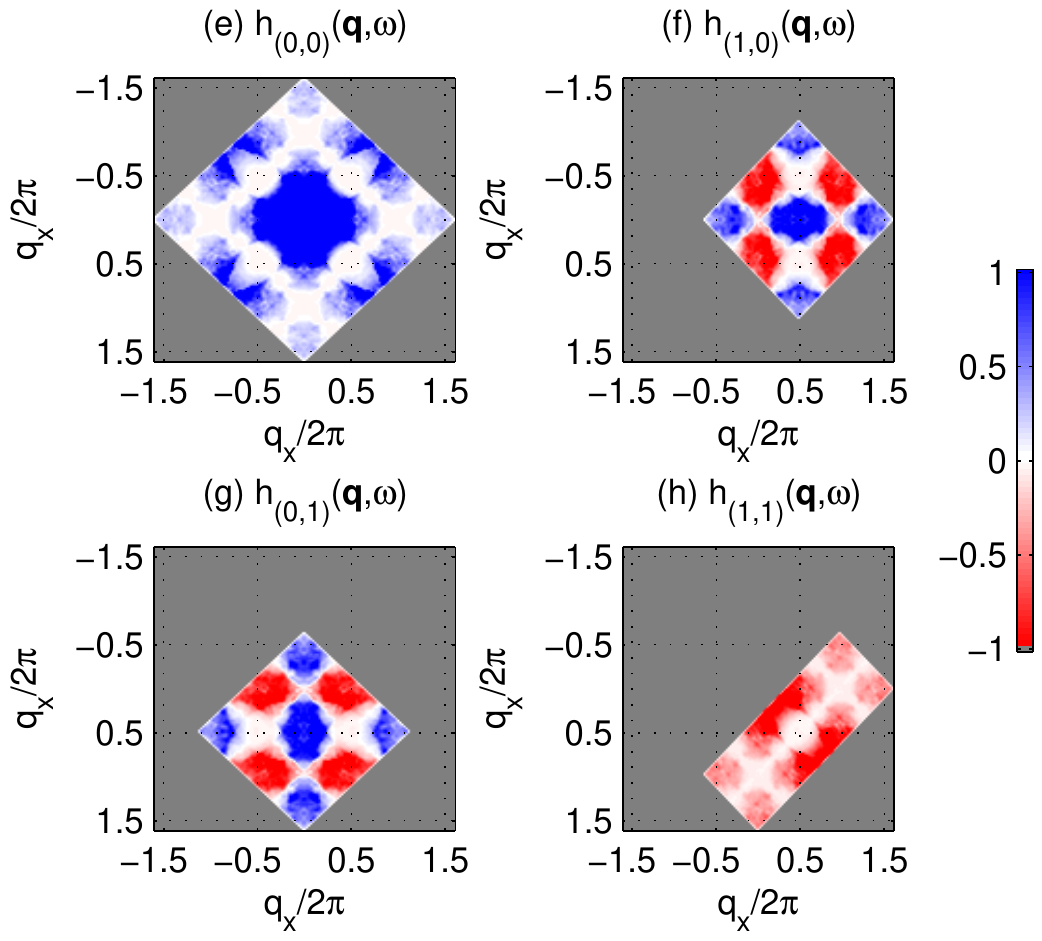}
\caption{{\bf Numerical holographic maps for alternative types of scatterers}. (a-d) Local modulation of the chemical potential. (e-h) Local modulation of the hopping on a single plaquette. These numerical maps were obtained using the same microscopic parameters as in Fig.~\ref{fig:dwave_th}, but with a different type of scatterer and are not consistent with the experimental data.}
\label{fig:dwave_v1}
\end{figure}

\newpage
\section{Nature of the local scatterer}
\label{app:scatterer}
The holographic maps presented in Fig.~\ref{fig:dwave_th} refer to a model of Friedel oscillations around a local modulation of the pairing gap. In this section we consider the effects of other types of modulations. Specifically, in Fig.~\ref{fig:dwave_v1}(a-d) we consider local modulations of the chemical potential, $T_k = \sigma_z$, where $\sigma_z$ is a Pauli matrix (see Methods section and Ref.~[\onlinecite{dallatorre15}] for details). The resulting holographic maps are inconsistent with the experimental findings, in particular because they are peaked at ${\bf q}_{\pi,\pi}=(0.5,0.5)$. This peak is a distinctive property of local modulations of the chemical potential\cite{dallatorre15} and is absent (or very weak) in the experimental data of Fig.~\ref{fig:dwave_exp}. In Fig.~\ref{fig:dwave_v1}(e-h) we consider a local modulation of the hopping on a plaquette $T_k = (\cos(k_x)+\cos(k_y))\sigma_z$. The resulting FT-STS maps do not display a star-shaped pattern around the Bragg peak (0,1) and are therefore again inconsistent with the experimental data.
We deduce that the experimentally-measured holographic maps shown in Fig.~\ref{fig:dwave_exp} are mainly due to local modulations of the pairing gap.
\newpage

\section{Energy dependence}
\label{app:energy}
\begin{figure}[b]
\includegraphics{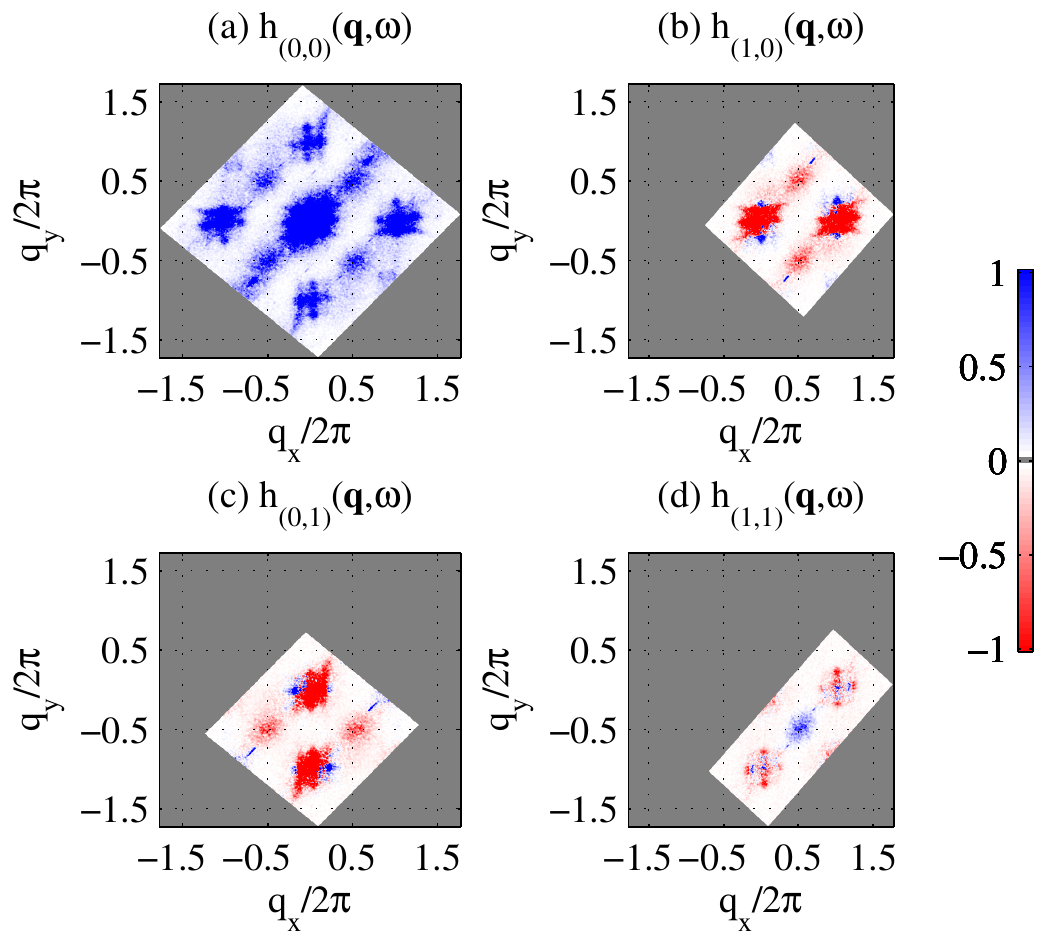}
\caption{{\bf Holographic maps at low voltage.} Same as Fig.~\ref{fig:dwave_exp} with $V=10$meV. The symmetry of these maps is less evident, probably due the coexistence of different types of impurities.}
\label{fig:dwave_10exp}
\end{figure}

The theoretical and experimental FT-STS maps shown in the text refer to an underdoped sample of Bi2201 at voltage $V = 40$meV. In this section we briefly discuss the energy dependence of the signal. Because by definition the Wannier functions are energy independent, our theoretical model predicts that the shape of the holographic maps should be roughly independent on the voltage. This prediction is confirmed by the experimental data that we considered, although at low voltages the $d$-wave symmetry of $h_{(1,0)}$ and $h_{(0,1)}$ is less evident (see Fig.~\ref{fig:dwave_10exp}). A possible explanation for this phenomenon is related to the coexistence of other types of scatterers, whose relative contribution to the scattering amplitude varies as a function of voltage. Indeed, the low-voltage experimental data of Fig.~\ref{fig:dwave_10exp} shows sharp peaks at $(0.5,0.5)$ which can be attributed to local modulations of the chemical potential (see \ref{app:scatterer}).

\begin{figure}[t]
\includegraphics[scale=0.8]{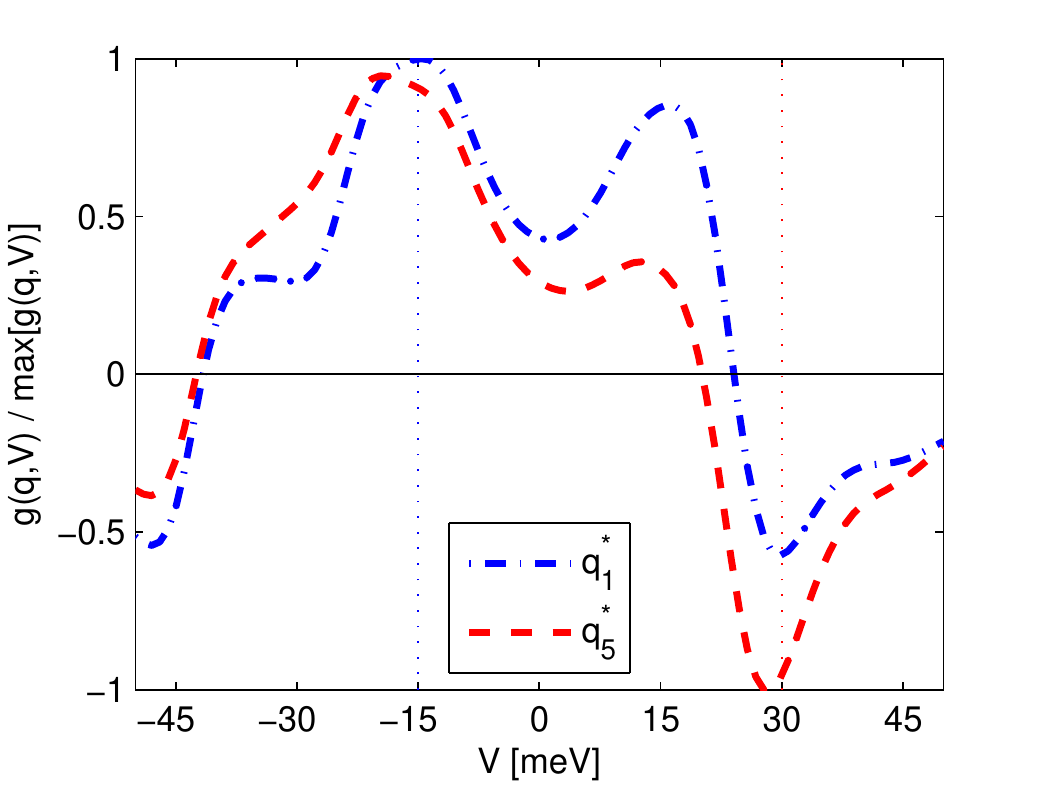}
\caption{{\bf Numerically-computed energy dependence of the FT-STS signal at wavevectors $q_1^*$ and $q_5^*$}. Each curve is renormalized independently.  We find that the $q_5^*$ signal is mostly anti-symmetric with respect to $V\to-V$ in agreement with the experimental findings of Ref.~[\onlinecite{hamidian2015atomic}]}
\label{fig:dwave_energy}
\end{figure}

To further highlight the non-trivial dependence of the scattering amplitude on the voltage (for a fixed type of impurity), in Fig.~\ref{fig:dwave_energy} we plot the amplitude of the numerically-computed FT-STS map at wavevectors $q_1^* = (\delta,0)$ and $q_5^* = (1-\delta,0)$, with $\delta=0.25$, as a function of voltage. We find that the these two wavevectors have maximal intensities respectively at $V=-15$meV and $V=30$meV. This finding is in qualitative agreement with the experimental measurements of Ref.~[\onlinecite{hamidian2015atomic}], who found the associated $s'$ and $d$-form factors to be peaked respectively at $20$meV and $90$meV. To better understand the relation between theory and experiment, we need to recall that Ref.~[\onlinecite{hamidian2015atomic}] analysed an underdoped sample of Bi2212 whose pairing gap is\cite{fujita14} of about $60$meV, while our theory has $\Delta_0=18$meV : we find that in both materials $q_5^*$ is peaked at about 1.5 times the pairing gap. We additionally notice that the amplitude of the $q_5^*$ signal at $\Delta=30$meV has opposite sign than the amplitude of the signal at $\Delta=-30$meV. This feature was experimentally observed by Ref.~[\onlinecite{hamidian2015atomic}] and interpreted as evidence of an {\it antisymmetric} d-form factor competing order. In our model it is instead a simple consequence of the coherence factors appearing in the scattering amplitude $G_{k,k'}(\omega)$.

\newpage

\section{Determination of the Wannier function} 
\label{app:wannier}
\begin{figure}[b]
\includegraphics[scale=0.9]{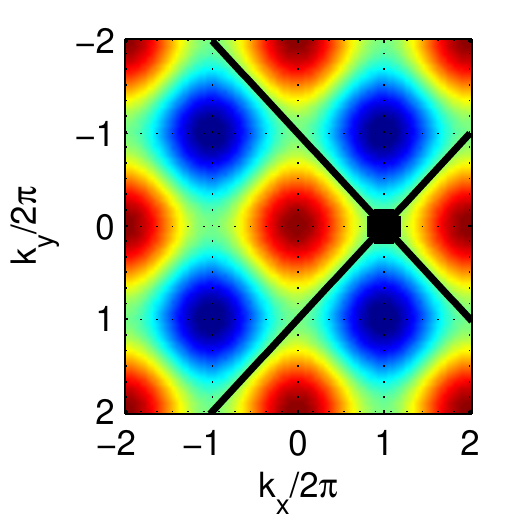}
\caption{{\bf Fourier transform of the $d$-wave Wannier function (\ref{eq:dWannier})}. The function is periodic over (2,0) and (0,2), and has a $d$-wave symmetry around the first Bragg peak (black dot).}
\label{fig:WFT}
\end{figure}

\vspace{-0.5cm}
The Wannier function inferred from the comparison between theoretical calculations and experimental observations is plotted in Fig.~\ref{fig:wf}(a) and has a $d$-wave shape. Interestingly, its four lobes are centred around the in-plane Oxygen sites. To understand the relation between this property of the Wannier function and the resulting $g$ maps it is useful to approximate $W(x,y)$ by the sum of four delta functions, centred respectively at the four Oxygen sites: 
\vspace{-0.3cm}
\be W(x,y) = \delta(x-a/2) + \delta(x+a/2) - \delta(y-a/2) - \delta(y+a/2)\;.\label{eq:dWannier} \ee
Its Fourier transform $W(k_x,k_y) = \cos(k_x a/2) - \cos(k_y a/2)$ is plotted in Fig.~\ref{fig:WFT} and is periodic over ${\bf k}=(2,0)$ and (0,2). An important consequence of this periodicity is that $W(k_x,k_y)$ vanishes along the lines ${\bf k} = (1-k,k)$ and ${\bf k} = (1+k,k)$ (black lines), leading to a d-wave shape around the first Bragg peak $(1,0)$ (black dot). The Wannier function enter into the calculation of FT-STS maps in a non-trivial way (see Eq.~(\ref{eq:FFG})), which however preserves the periodicity of the signal. In the case of Eq.(\ref{eq:dWannier}) the resulting $g$ map is therefore expected to be periodic over (2,0) and (0,2), and to vanish along the lines ${\bf q}=(1-q,q)$ and ${\bf q}=(1+q,q)$. Actual Wannier functions have a finite extension and are not exactly periodic in momentum space. The $d$-wave shape of the g-maps in the first Brillouin zones is however preserved. 

To better understand the role of the Wannier function, in Figs.~\ref{fig:theory1}-\ref{fig:theory4} we plot $g$ maps and $h$ maps for different trial functions, with $s$-wave and $d$-wave symmetry, and different extensions. As explained in the respective captions, these Wannier functions do not reproduce the experimental results of Fig.~\ref{fig:dwave_exp}, with the exception of Fig.~\ref{fig:theory3}. The Wannier function used in this figure equals to the absolute value of Eq.~(\ref{eq:Wx}) and leads to similar holographic maps.

\linespread{1}

\def \commoncaption {{ (a) Real-space Wannier function (b) $g$ map; (c) $h$ maps. In contrast to the results presented in the main text (Figs. \ref{fig:dwave_th} and \ref{fig:schematic-gmap}), the present choice of the Wannier function does not correctly reproduce the experimentally observed $h$ maps (Fig.~\ref{fig:dwave_exp})}}

\newcounter{i}
\forloop{i}{1}{\value{i}<5}
{
\begin{figure*}[h!]
\centering
\begin{tabular}{c c c}
\begin{tabular}{c}
(a) W(x,y)\\
\includegraphics[scale=0.7]{wf_th\arabic{i}} \\
\end{tabular}&
\begin{tabular}{c}
(b) $g({\bf q},\omega)$\\
\includegraphics[scale=0.7]{gmap_th\arabic{i}}
\end{tabular}
&
\begin{tabular}{c}
(c)\\
\includegraphics[scale=0.7]{dwave_th\arabic{i}}
\end{tabular}
\end{tabular}
\ifthenelse{\value{i}=1}{\caption{Theoretical calculations for a s-wave Wannier function $W({\bf x})=e^{-(x^2+y^2)/(2\sigma^2)}$ $\sigma_x=0.5a$. This wavefunction was used by Ref.~\onlinecite{dallatorre13}. \commoncaption. In particular, the present theoretical predictions for $G_0=(1,0)$ indicate a single broad peak around ${\bf q}=0$ while the experimental observations display for distinct peaks at ${\bf q_0}\pm(0.25,0)$ and ${\bf q_0}\pm(0,0.25)$.   }}{}
\ifthenelse{\value{i}=2}{\caption{Theoretical calculations for an extended $s$-wave Wannier function $W({\bf x})=(x^2+y^2)~e^{-(x^2+y^2)/(2\sigma^2)}$ with $\sigma_x=0.3a$.\commoncaption. In particular, the four distinct peaks observed at $G_0=(1,0)$ and ${\bf q_0}\pm(0.25,0)$ and ${\bf q_0}\pm(0,0.25)$ have identical sign, while in the experiment they have alternating signs.}}{}
\ifthenelse{\value{i}=3}{\caption{Theoretical calculations for an extended $s$-wave Wannier function $W({\bf x})=| x^2-y^2 |~ e^{-(x^2+y^2)/(2\sigma^2)}$ with $\sigma_x=0.3a$. It corresponds to the absolute value of the wavefunction considered in main text and leads to similar $h$ maps. This example clarifies that FT-STM maps are sensitive to the absolute value of the Wannier functions and not to their phase.}
\label{fig:theory\arabic{i}}}{}
\ifthenelse{\value{i}=4}{\caption{Theoretical calculations for a $d$-wave Wannier function $W({\bf x})=(x^2-y^2)~e^{-(x^2+y^2)/(2\sigma^2)}$ with $\sigma_x=0.7a$. This wavefunction is similar to the one used in the main text, but with a larger width, selected to closely reproduce the first-principle calculations of Ref.~\onlinecite{kreisel15}. \commoncaption. Interestingly the present results are analogous to the one obtained for an $s$-wave Wannier function in Fig.~\ref{fig:theory1}}}{}
\label{fig:theory\arabic{i}}
\end{figure*}
}

\newpage

\end{document}